\documentclass[prb,aps,showpacs,twocolumn,unsortedaddress,superscriptaddress]{revtex4-1}
\usepackage{bm,psfrag,amsmath,amssymb,epsfig,grffile,float}
\usepackage{tabularx,graphicx,epsf,color}
\usepackage[usenames,dvipsnames,svgnames,table]{xcolor}
\usepackage{booktabs}

\def\sss#1{{\scriptscriptstyle #1}}
\def\ssr#1{{\sss{\rm #1}}}
\def\frac#1#2{{\textstyle{#1 \over #2}}}
\def\half{\frac{1}{2}}

\def\nd{^{\vphantom{\dagger}}}
\def\ns{^{\vphantom{*}}}
\def\yd{^{\dagger}}

\def\eg{{\it e.g.\/}}

\def\ve{\varepsilon}

\def\SK{{\textsf K}}
\def\Rf{{\rm f}}
\def\Nf{N\ns_{\Rf}}

\def\Bk{\boldsymbol{k}}
\def\BR{\boldsymbol{R}}
\def\Bsigma{\boldsymbol{\sigma}}
\def\CE{{\mathcal E}}

\def\RTheta{{\rm\Theta}}

\def\vphi{{\varphi}}

\def\vF{v\nd_\ssr{F}}
\def\vei{\ve\ns_{\rm imp}}
\def\CEi{\CE\ns_{\rm imp}}
\def\mui{\mu\ns_{\rm imp}}
\def\muTI{\mu\ns_{\rm TI}}

\begin{document}

\title{Dirac node engineering and flat bands in doped Dirac materials}
\author{Anna Pertsova}
\affiliation{Nordita, Roslagstullsbacken 23, SE-106 91 Stockholm, Sweden}
\author{Peter Johnson}
\affiliation{Brookhaven National Laboratory, Condensed Matter Physics Materials Science Department, Upton, NY 11973-5000, USA}

\author{Daniel P. Arovas}
\affiliation{Department of Physics, University of California, San Diego CA, USA}

\author{Alexander V. Balatsky}
\affiliation{Nordita, Roslagstullsbacken 23, SE-106 91 Stockholm, Sweden}
\affiliation{Deptment of Physics, University of Connecticut,  Storrs, CT 06269, USA}

\date{today}

\begin{abstract}
We suggest the tried approach of impurity band engineering to produce flat bands and additional nodes in Dirac materials. We show that surface impurities give rise to nearly flat impurity bands close to the Dirac point. The hybridization of the Dirac nodal state induces the splitting of the surface Dirac nodes and  the appearance of new nodes at high-symmetry  points of the Brillouin zone. The results are robust and not model dependent: the tight-binding calculations are  supported by a low-energy effective model of a topological insulator  surface state hybridized with an impurity band.  Finally, we address the effects of electron-electron interactions between localized electrons on the impurity site. We confirm that the correlation effects, while producing  band hybridization and Kondo effect, keep the hybridized band flat. Our findings open up prospects for impurity band engineering of nodal structures and flat-band correlated phases in doped Dirac materials.   
\end{abstract}

\maketitle
\section{Introduction}
Impurity band engineering is at the core of modern semiconducting industry where impurity bands enable functionality of a semiconductor. Similarly, electronic structure and topology of Dirac materials \cite{Wheling_adv_in_phys2014} (DMs) can be manipulated by impurity doping. A well known example of such manipulation is the quantum anomalous Hall effect (QAHE)\cite{Yu2010,Chang2013},  which is a new 
quantum state of matter observed in magnetically-doped three-dimensional (3D) topological insulators (TIs). 
It occurs as a result of a gap opening at the Dirac node of TIs due to broken time-reversal symmetry. 
It is known that impurities give rise to low-energy resonant states near Dirac nodes~\cite{Biswas_prb2010,Annika_prb2012,
Annika_prb2012_2}. For a magnetically-doped 3D TI, 
the magnetic energy gap is filled with impurity resonant states. Hence, disorder effects have significant implications 
for QAHE~\cite{Annika2015,Sessi2016}. 
In this work, we consider another example of impurity band engineering in  DMs. Specifically, we propose to use impurity bands  to introduce flat bands and produce additional Dirac nodes in DMs. 

Flat bands can occur in a variety of systems, including electronic materials such as superconducting wire networks and engineered 2D atomic lattices, 
in optical lattices of cold atoms, and in photonic systems such as waveguide arrays and exciton-polariton condensates~\cite{Leykam2018}. 
Due to the 
quenching of kinetic energy, flat bands are highly susceptible to interactions. In particular, flat electronic bands are expected to 
give rise to interaction-driven quantum phases, such as superconductivity and Bose-Einstein condensation.  Nearly flat electronic bands 
can be found in “slow” DMs, in which Dirac states exhibit extremely small velocity, resulting in a large coupling constant~\cite{Triola_prb2015}.  
Recently, there has been a renewed interest in flat bands due to the discovery of superconductivity in twisted bilayer graphene near 
the so-called  magic angles which host “slow” Dirac fermions close to the charge-neutrality  point~\cite{Bistritzer_pnas2011, Cao2018_part2,Cao2018}. 

Here we consider impurity-induced flat bands that emerge in doped DMs, \eg\ in 3D TIs, with periodically-arranged impurities. In a certain range of impurity potentials, 
these impurity bands appear near the Dirac nodes and hybridize with the Dirac states.

We demonstrate, by using an effective model and tight-binding calculations for a typical 3D TI with a single Dirac node at the $\Gamma$ point, that when the impurity resonance state is energetically close to the Dirac node, such hybridization results in the splitting of the original Dirac node at $\Gamma$ and in the appearance of additional nodes at other high-symmetry points in the Brillouin zone. 

We also show, using a slave boson approach, how a Dirac conduction band hybridizing 
with a strongly interacting localized band via a Kondo coupling $J$ also leads to 
this physics, with a strongly renormalized chemical potential for the local orbitals,
provided that $J$ is sufficiently large (on the order of the conduction electron
bandwidth).

\section{Model} 
Our theoretical modelling is based on the sp$^{3}$ Slater-Koster 
tight-binding (TB) model 
for Bi$_2$Se$_3$ 3D TI, with parameters fitted to ab initio  calculations obtained
with Wien2k package~\cite{Kobayashi2011,Pertsova2014}. 
The TB model for pristine Bi$_2$Se$_3$
includes $s$ and $p$ orbitals and Slater-Koster hopping elements between atoms 
in the same atomic layer and between atoms in first and second nearest-neighbor
layers. The spin-orbit interaction is incorporated in the intra-atomic matrix elements.

For surface calculations with impurity doping on the (111) surface, we consider a slab consisting of five quintuple layers (QLs) and a $3\times{3}$ surface supercell 
\cite{Pertsova2014} (unless stated otherwise). 
An impurity substitutes an atom in the topmost Bi layer  
and is described by a local on-site potential 
\cite{Biswas_prb2010,Annika_prb2012, Annika_prb2012_2}.
A point-like impurity potential $U$ acts 
as a uniform shift to the on-site energy of the impurity atom. The impurity potential  introduces localized 
impurity states that can affect electronic states in the vicinity of the Dirac node 
\cite{Annika_prb2012}. Microscopic tight-binding model calculations are accompanied by low-energy continuum model calculations for a TI surface state hybridized with a doubly degenerate impurity band. 

\section{Results}
\subsection{Tight-binding calculation for 3D TI}
\begin{figure*}[ht]
\centering
\includegraphics[width=0.98\linewidth,clip=true]{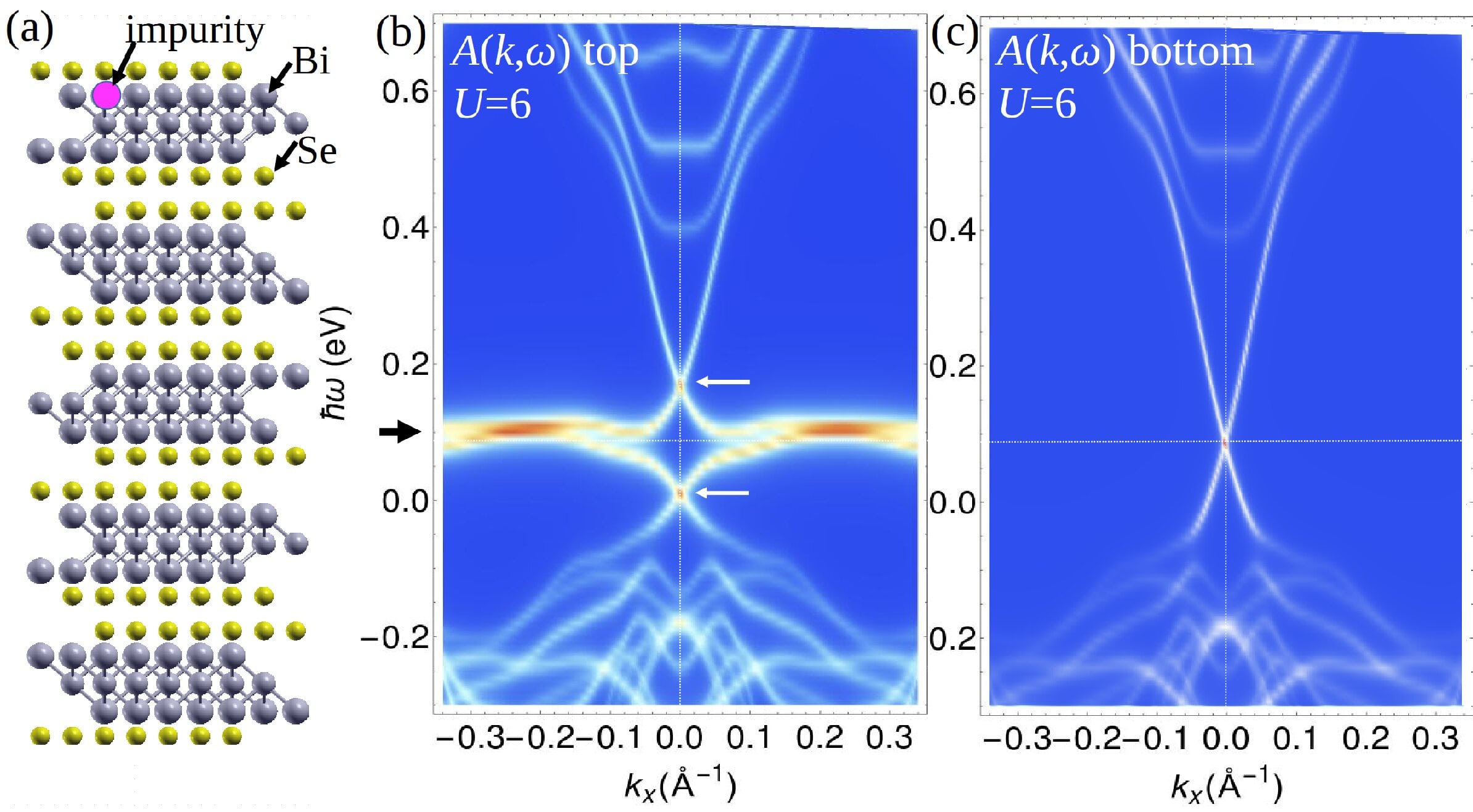}
\caption{Impurity-engineered flat bands in a 3D TI. (a) A supercell of a 5QL thin film of Bi$_2$Se$_3$ 3D 
TI with an impurity atom substituting Bi in the topmost Bi atomic layer. The size of the surface supercell is $3\times{3}$,  
corresponding to 11$\%$ surface doping. The impurity potential is $U=6$~eV. The calculated atomic-layer resolved  spectral 
function, $A(k,\omega)$, (b) on the top surface containing the impurity and (c) on the bottom (undoped) surface. A black arrow in (b) shows the position 
of the impurity band. White arrows in (b) indicate the modified surface Dirac points. Thin dashed 
lines show the position of the Dirac point in undoped film ($U=0$).}
\label{fig_flat_bands1}
\end{figure*}

\begin{figure*}[ht!]
\centering
\includegraphics[width=0.90\linewidth,clip=true]{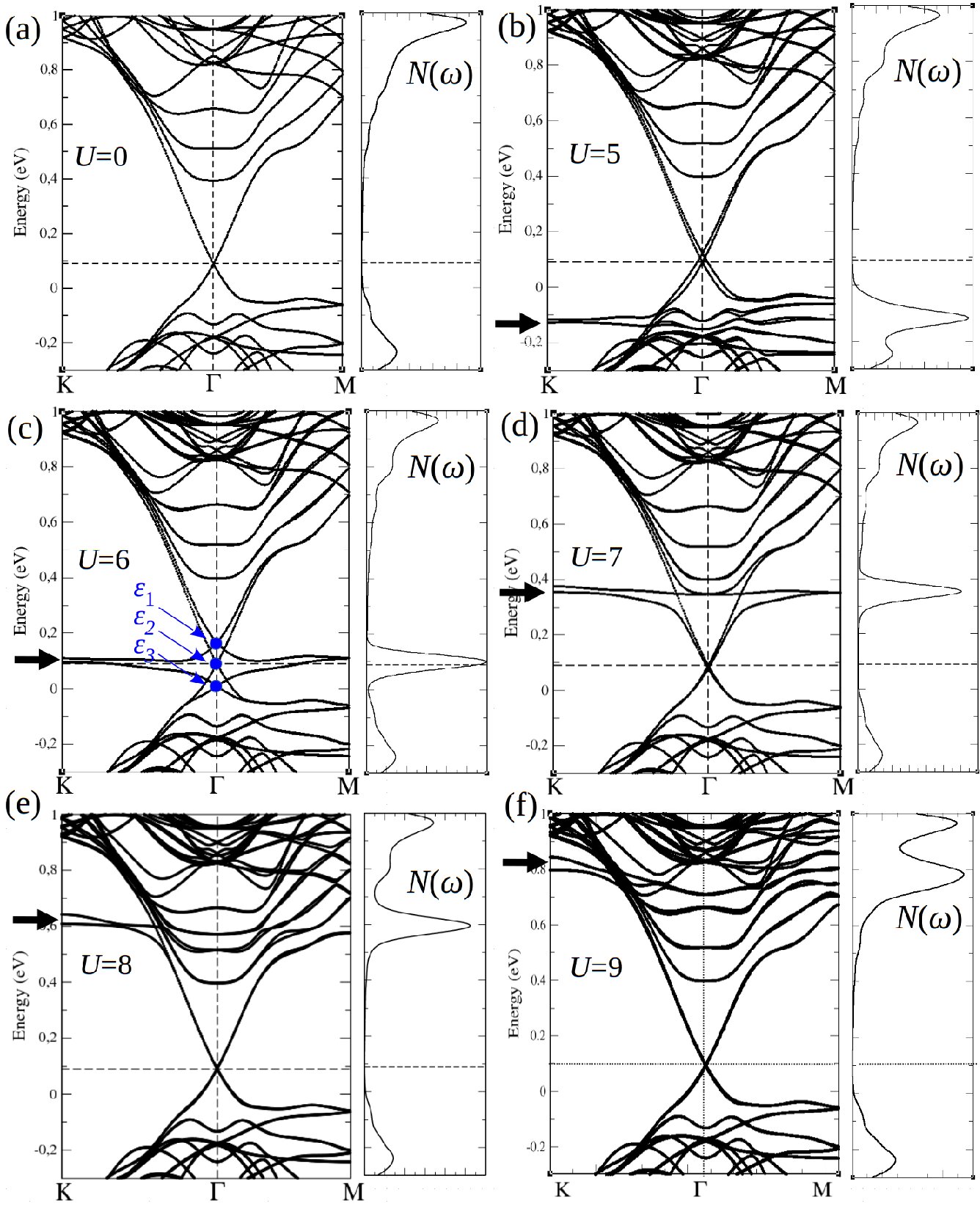}
\caption{The 
calculated band structures and DOS for several values of $U$ ($U=0-9$~eV). Black arrows show the position 
of the impurity band. Thin dashed 
lines show the position of the Dirac point in the undoped film ($U=0$).}
\label{fig_flat_bands2}
\end{figure*}

The results of a 
representative TB calculation with $U=6$~eV  are shown in Fig.~\ref{fig_flat_bands1}. The supercell structure with a substitutional impurity near the top surface is shown in  Fig.~\ref{fig_flat_bands1}(a). The spectral  function corresponding to atomic layers along the growth direction of the TI slab is calculated by averaging the atom-resolved spectral function over the atoms in a specific layer. The spectral functions of the top and bottom surfaces are shown in  Fig.~\ref{fig_flat_bands1}(b) and (c), respectively. The presence of surface impurities leads to impurity 
resonance states 
that appear as 
nearly flat bands in the spectral function of the doped surface [Fig.~\ref{fig_flat_bands1}(b)], while the undoped surface exhibits unperturbed Dirac states.   

Figure~\ref{fig_flat_bands2} shows the calculated band structures and the density of states (DOS) for a range of impurity potentials $U$. For $U=0$, the Dirac states of the top and bottom surfaces of the slab are degenerate. With increasing $U$, a nearly dispersionless band emerges in the valence band, and the degeneracy between the top and bottom Dirac states is lifted. The Dirac node of the undoped surface remains pinned at the position of the Dirac node of the pristine system, while the Dirac node of the doped surface is shifted in energy. With increasing the impurity potential further, the impurity band crosses the Dirac node of the top surface and shifts further up in energy. For large $U$, the impurity band merges with the conduction band and the degeneracy of the top and bottom Dirac states is restored. 

\begin{figure}[ht]
\centering
\includegraphics[width=0.98\linewidth,clip=true]{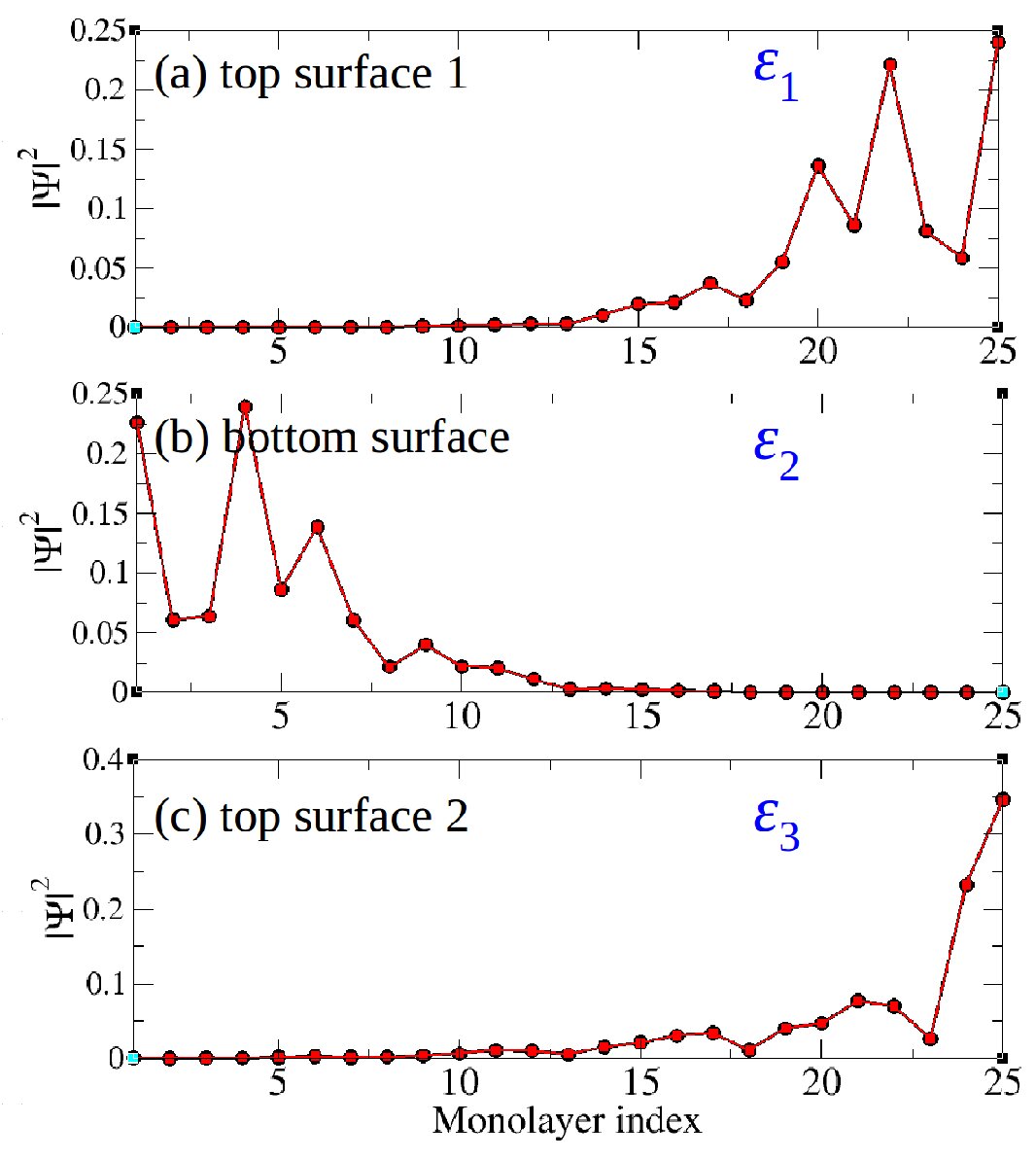}
\caption{The absolute value of the wavefunction of the three doubly degenerate states $\ve_j$ ($j=1,2,3$) at 
$\Gamma$, 
marked in Fig.~\ref{fig_flat_bands2}(c), as a function of atomic position along the slab. 
 Atomic index $i=1(N)$ corresponds to bottom(top) surface. The states $\ve_1$ and 
 $\ve_2$ are the two Dirac nodes of the top surface, split from the original node by the  impurity resonance. The impurity potential is $U=6$~eV.}
\label{fig_flat_bands3}
\end{figure}

\begin{figure*}[ht]
\centering
\includegraphics[width=0.98\linewidth,clip=true]{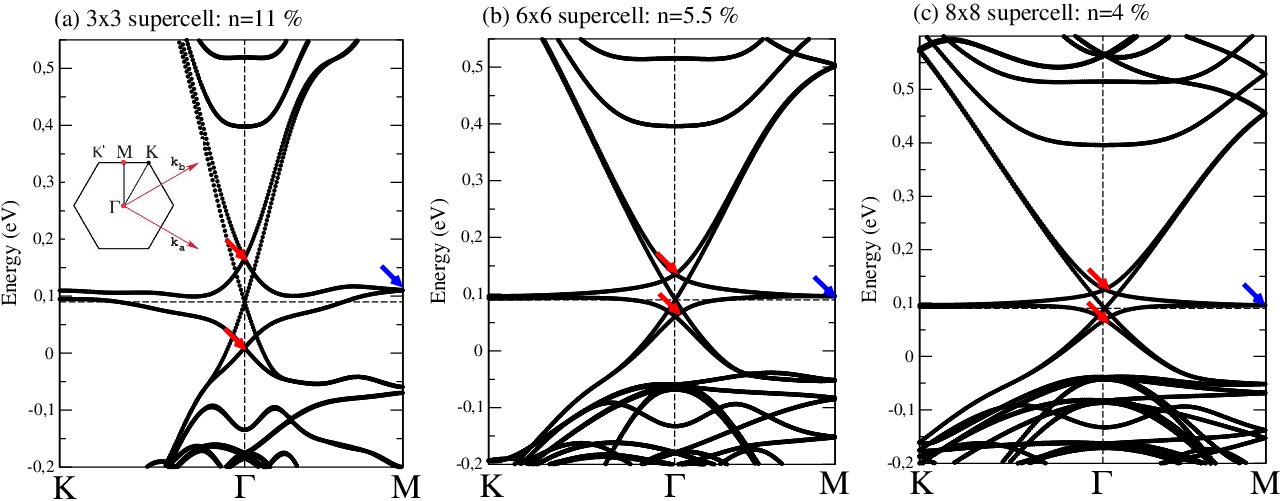}
\caption{The calculated band structures with  $U=6$~eV for decreasing concentration of 
surface impurities: (a) $n=11\%$, (b) $n=5.5\%$, and (c) $n=4\%$. The inset in (a) shows the surface Brillouin zone.}
\label{fig_concentr}
\end{figure*}

The impurity states also appear as peaks in the DOS as shown in the right hand-side panels in Fig.~\ref{fig_flat_bands2}. Their position is controlled by the non-magnetic impurity potential $U$ and coincides with the position of the flat impurity band. 


When the impurity resonance state falls in the vicinity of the node, which corresponds to $U=6$~eV in this model [Fig.~\ref{fig_flat_bands2}(c)], the Dirac spectrum is drastically modified.  The Dirac 
node of the doped surface is split into two nodes, displaced vertically in energy,  and the surface states hybridize with the flat impurity band. Figure~\ref{fig_flat_bands3} shows the spatial distribution 
of the wavefunctions of the three nodes at $\Gamma$, which are marked as $\ve_i$ in  Fig.~\ref{fig_flat_bands2}(c) and correspond to the Dirac nodes of the bottom surface and the split nodes of the top surface. 
The two new nodes at $\Gamma$ are predominantly localized at the top surface. However, they are not pure surface states 
and are hybridized with the impurity band localized on the impurity site.

This feature was noticed in previous theoretical work on 3D TIs, which used the scattering matrix approach with and without disorder to study impurity resonances~\cite{Annika2015,Zhong2017}. In contrast to our tight-binding calculations, this approach does not reply on the use of a supercell. The splitting of the Dirac node due to  coherent impurity scattering and the appearance of dispersionless band at zero energy was shown in graphene in the presence of vacancies~\cite{Zhu_prb2012}. 
 A similar effect was found in ab initio calculations of TI/normal semiconductor heterostructures, where a semiconductor valence band  hybridizes with TI surface states~\cite{Seixas_natcomm2015}. This result was qualitatively explained by an effective model, based on the low-energy surface state Hamiltonian hybridized with a trivial band from a proximal semiconductor layer.  More generally, the possibility of reshaping the topologically protected surface states with localized impurity resonances was demonstrated in \cite{Xu_natcomm2017} using numerical simulations and scanning tunneling microscopy (STM).
 
Although the splitting of the Dirac nodes in the presence of disorder was noticed in previous work~\cite{Annika2015, Zhong2017}, it has not been studied in the context of topology and impurity-controlled nodal structure of DMs. To further illustrate the details of this effect, we show in Fig.~\ref{fig_concentr} the calculated band structures for $U=6$~eV and for different surface impurity concentration. As one can see from Fig.~\ref{fig_concentr}(a), the Dirac node of the top surface is split vertically into two nodes (red arrows) and is hybridized with the impurity band, while the bottom surface states remain unaffected by doping. The splitting between the two nodes at $\Gamma$ decreases with increasing the doping concentration. 

An additional doubly degenerate state appears at the Brillouin zone corner ($M$ points) and is marked by a blue arrow in Fig.~\ref{fig_concentr}. Here, we consider non-magnetic doping, hence time-reversal symmetry guarantees Kramers degeneracy at the  time-reversal invariant momenta $\Gamma$ and $M$.  The surface of Bi$_2$Se$_3$ 3D TI is a triangular lattice. Due to the symmetry of  the corresponding hexagonal Brillouin zone [see the inset in Fig.~\ref{fig_concentr}(a)], there are three non-equivalent $M$ points each hosting a doubly degenerate state. 

The splitting and generation of new nodes leads to a natural question of whether the topology of the surface states is somehow affected by non-magnetic impurities. As confirmed by the calculations, the nodes remain gapless; however, their number and position change. We verified that the topological properties are preserved in the presence of non-magnetic doping despite the modified nodal structure. The Dirac states in a 3D TI are characterized by spin-momentum locking in the vicinity of the node.  We define the helicity $h$ as the eigenvalue of the helicity operator $\Bsigma\cdot{\hat\Bk}$, where ${\hat\Bk}=\Bk/|\Bk|$. This can be visualized as the direction of rotation of the spin of an energy eigenstate as the momentum changes clockwise from $+k_y$ to $-k_y$. 
We found numerically that the helicity of the doped surface, calculated by summing the helicities of the nodes at $\Gamma$ and $M$, is the same as the helicity of the pristine surface. Thus, the peculiar splitting of the surface state node by impurity resonance necessitates the appearance of new nodes at the Brillouin zone corners to preserve the topology.  

\subsection{Analytical model of a Dirac spectrum hybridized with an impurity band}
We will illustrate the splitting of the Dirac node by hybridization 
with an impurity band by using an effective low-energy model. We consider the following Hamiltonian
\begin{equation}\label{Dirac}
   H=
  \left( {\begin{array}{cc}
   H\ns_{\mathrm{TI}} & V \\
   V\yd & \CEi \\
  \end{array} } \right), 
\end{equation}
where 
\begin{equation}
   H\ns_\mathrm{TI}=
  \left( {\begin{array}{cc}
  -\muTI & \vF\,(k_x-i k_y) \\
   \vF\,(k_x+i k_y) & -\muTI \\
  \end{array} } \right). 
\end{equation}
is the Hamiltonian of a 3D TI surface (or graphene); $v_F$ is the  Fermi velocity. Here $V=\gamma\hat{I}$ is the coupling matrix, $\gamma$ is the coupling strength, $\hat{I}$ is a $2\times{2}$ identity matrix. $\CEi=\vei(\Bk)\hat{I}$ is the impurity band Hamiltonian, where $\vei(\Bk)$ is the impurity band dispersion. 
We consider $\vei(\Bk)=
{a}\Bk^2-\mui$; the case $a\ne{0}$ corresponds to a quadratic band, 
 while $a=0$ gives a flat band. The resulting bandstructures are shown in Fig.~ 5.

\begin{figure*}[!t]
\begin{center}
\centering
\includegraphics[width=0.95\linewidth,clip=true]{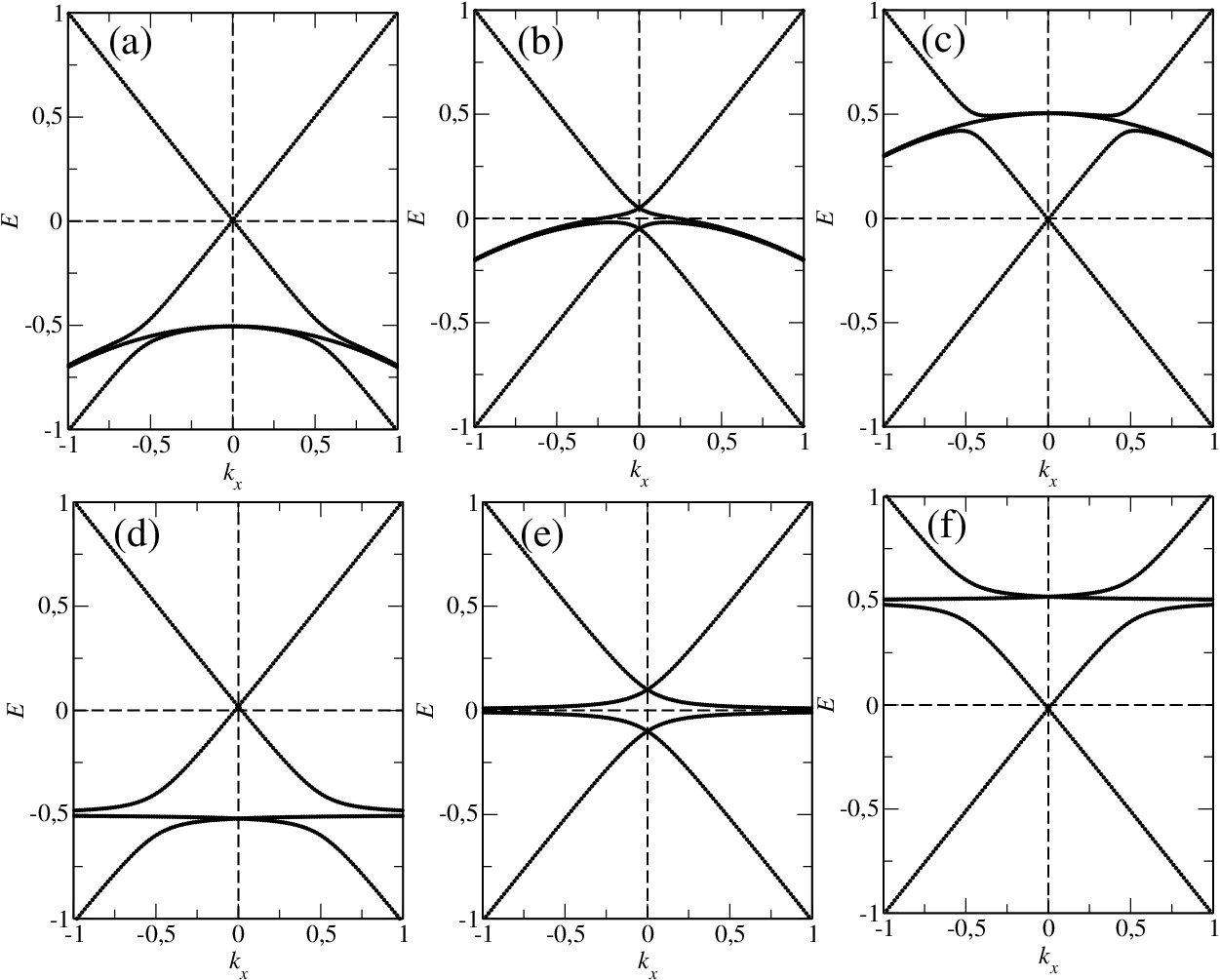}
\caption{The calculated bands of the effective low-energy  model for $\muTI=0$ and (a,d) $\mui=0.5$, (b,e) $\mui=0.0$, and (c,f) $\mui=-0.5$. Top panels are for a quadratic band, bottom panels are for a flat impurity band.}
\end{center}
\label{fig_flat_bands5}
\end{figure*}

The eigenvalues of the Hamiltonian in Eq.~(\ref{Dirac}) can be found from the following equation
\begin{equation}
    (E^2-\vei(\Bk)\,E-\gamma^2)^2=(E-\vei(\Bk))^2\,(\vF\Bk)^2.
\end{equation}
The pair of doubly degenerate states at $\Gamma$  are given by the solution at  $k_x=k_y=0$ 
\begin{equation}
    E\ns_{1,2}=-\half\mui\pm\half\sqrt{\mu_{\rm imp}^2+4\gamma^2}.
\end{equation}
For $\mui\gg \gamma$ ($\gamma\ne{0}$), we have $E\ns_1=0$ and $E\ns_2=\mu_{\rm imp}$.

\subsection{Hybridization via Kondo coupling}
For completeness we also analyze the case of correlated impurity bands. While the presence of the flat bands due to impurities is natural, the question can be asked about stability of our results in the presence of correlations. To address this, we now consider an interacting model of a Kondo lattice. We use an analysis based on a localized, $\textsf{SU}(2)\times\textsf{SU}(N)$-degenerate $f$-band of electrons coupled to
a conduction electron band with Dirac cones.  The Hamiltonian is
\begin{align}
H&=\sum_{\Bk,m\atop\sigma,\sigma'} h\nd_{\sigma\sigma'}(\Bk)\,c\yd_{\Bk m\sigma}\,c\nd_{\Bk m\sigma'} + \ve\ns_0\sum_{\BR,m,\sigma} f\yd_{\BR m\sigma}\,f\nd_{\BR m\sigma}\nonumber\\
&\quad -{J\over\Nf}\,\sum_{\BR}\sum_{m,m'\atop\sigma,\sigma'} \colon 
c\yd_{\BR m\sigma} \,f\nd_{\BR m\sigma}\,
f\yd_{\BR m'\sigma'}\,c\nd_{\BR m'\sigma'} \colon\quad.
\end{align}
where $\sigma$ is an $\textsf{SU}(2)$ index.  which could label different sublattices
in a graphene or flux phase~\cite{FluxPhase88} structure.  The index 
$m\in\{1,\ldots,\Nf\}$ labels the flavor, and we shall be interested in the limit
$\Nf\to\infty$.  The $f$-electrons are presumed to be strongly interacting, satisfying
the constraint $\sum_{m,\sigma} f\yd_{\BR m\sigma} f\nd_{\BR m\sigma}= q\Nf$,
where $q$ is the fixed filling fraction of the flavor orbitals in each unit cell,
which is conserved by $H$.  This constraint is enforced by Lagrange multipliers
$\lambda\ns_{\BR}$ at each site.  Invoking the well-established mean field treatment~\cite{Hewson1993}, the quartic term is decoupled via a Hubbard-Stratonovich
transformation with complex local fields $V\ns_{\BR}$, and the Kondo term becomes
\begin{align}
H\ns_{\SK}&=\sum_{\BR} {\Nf\,|V\ns_{\BR}|^2\over J} + \sum_{\BR,m,\sigma}
\big( V\ns_{\BR}\,c\yd_{\BR m\sigma}\,f\nd_{\BR m\sigma}\\
&\hskip 1.8in + V^*_{\BR} \,f\yd_{\BR m\sigma}\,c\nd_{\BR m\sigma}\big)\quad.\nonumber
\end{align}
Assuming a mean field solution where $\lambda\ns_{\BR}=\lambda$ and $V\ns_{\BR}=V$ are 
spatially uniform, the mean field Hamiltonian becomes
\begin{align}
H^\textsf{MF} &= N\Nf\bigg({|V|^2\over J} - q\lambda\bigg)+ \\
&\hskip-0.3in \sum_{\Bk,m\atop\sigma,\sigma'} 
\begin{pmatrix} c\yd_{\Bk m\sigma} & f\yd_{\Bk m\sigma} \end{pmatrix}
\begin{pmatrix} h\ns_{\sigma\sigma'}(\Bk) & V\,\delta\ns_{\sigma\sigma'} \\ V^*\,\delta\ns_{\sigma\sigma'} & (\ve\ns_0+\lambda)\,\delta\ns_{\sigma\sigma'}\end{pmatrix}
\begin{pmatrix} c\nd_{\Bk m\sigma'} \\ f\nd_{\Bk m\sigma'} \end{pmatrix}\ .\nonumber
\end{align}
At this point, we can work in the diagonal basis of $h\ns_{\sigma\sigma'}(\Bk)$, 
whose eigenvalues $E\ns_\alpha(\Bk)$ are upper and lower bands which touch at certain
$\Bk$ values where there are Dirac cones.  We also define $\ve\equiv\ve\ns_0+\lambda$, which is the renormalized $f$-level energy. For  simplicity of calculation, we adopt
a model of the $c$-electron density of states $g(E)$, with 
$g(E)=2\,|E|\, W^{-2}\,\RTheta\big(W-|E|\big)$.  At $T=0$, the dimensionless free
energy per site per flavor, in units of the $c$-electron half-bandwidth $W$, is
\begin{equation}
\vphi=W^{-1}\,\CE\big(|V|^2,\ve\big) + {2\over W^3}\!\int\limits_{-W}^{E^*}\!\! dE\,
|E|\,\xi\ns_{-}(E)\quad,
\end{equation}
where
\begin{equation}
\CE\big(|V|^2,\ve\big)={|V|^2\over J} - q(\ve-\ve\ns_0)
\end{equation}
is the non-fermionic contribution to the energy, 
\begin{equation}
\xi\ns_\pm(E)=\half (E+\ve )\pm \half \sqrt{(E-\ve)^2 + 4\,|V|^2}  
\end{equation}
are the energies of the hybridized $c$-$f$ bands (upper and lower), and $E^*$ is
defined by $\xi\ns_{-}(E^*)=0$, where we position our Fermi level.

The mean field solution is obtained by extremizing $\vphi$ with respect to the
parameters $|V|$ and $\ve$.  We defer the description of the complete solution
to a future publication and describe here some limiting results.  Writing 
$\ve\equiv x W$, $|V|\equiv \sqrt{r}\,\ve$, we obtain a solution
to the mean field equations only when $J>J\ns_{\rm c}=\half W$.  Defining
$\zeta\equiv 1-{W\over 2J}$, in the limit where $0<\zeta\ll 1$ we obtain the equations
\begin{equation}
x\ln x^{-1}=\zeta\qquad,\qquad r = {q\over 2\zeta\,x(\zeta)}\quad,
\end{equation}
where $x(\zeta)$ is a solution of the first of these mean field equations.
That a critical value of $J$ on the order of the bandwidth is necessary in order
to obtain a solution is expected from the work of Fradkin and others on magnetic
impurities in Dirac systems\cite{PhysRevLett.64.1835,PhysRevB.53.15079}.  In the 
present context, we find then that for $J>J\ns_{\rm c}$, strong interaction physics
within a localized $f$-band nominally located below the Fermi level results in a
renormalization of the $f$-electron energy, pushing it up to just above the chemical
potential, where the $f$-band can effectively hybridize with the Dirac $c$-band
and yield the nearly flat bands discussed previously.

\section{Discussion and Conclusions}
Although it can be problematic to achieve a regular lattice of impurities on a TI surface, there is a strong evidence for 
impurity-induced states in typical doped TI samples~\cite{Islam_prb2018}. Moreover, recent ARPES studies suggested that overlapping of impurity 
resonances with the Dirac node, a situation depicted  in Fig. 1(b), could be achieved in the experiment as the binding energy 
of the node changes with increasing the film thickness~\cite{Yilmaz_2020}. It is also possible to resolve, at least partially, the impurity 
bands by adjusting the photon energy of the pulse. Another promising platform that can be used to study impurity flat bands are artificially grown impurity superlattices on graphene. Such lattices can be realized by self-assembly of organic molecules on graphene deposited on the substrate~\cite{Jarvinen_nanolett2013}.

In Fig.~\ref{fig_flat_bands1}(b), we considered a situation where material parameters are such  
that the flat bands occur exactly 
at the Dirac point of the pristine material. Assuming that the chemical potential is at the Dirac node, this is the most favorable situation which allows access to flat bands.  In the new-generation of 3D TIs with composition (Bi,Sb)$_2$B$_3$ (B=Se,Te), chemical potential can be positioned within $10$~meV from the Dirac point. The location of the impurity resonances varies greatly with the type of material and dopant~\cite{Islam_prb2018}. 
Recent experiments indicate that favourable conditions can be achieved at least for some samples and dopants, \eg\ for  Cr in Bi$_2$Se$_3$ above the magnetic ordering temperature~\cite{Yilmaz_2020}.

However, 
in the majority of materials, the above conditions will not be satisfied. Further complications arise from the fact that impurity doping itself may change the position of the Fermi level.  For such situations,  optical pumping can be used to populate the flat bands. 
 Superconducting or excitonic pairing may occur between carriers residing in the flat bands and will have a transient nature~\cite{Triola_prb2017,Pertsova_prb2018}.
These ideas can be generalized to pumping of flat, or nearly-flat bands, which do not necessarily originate from impurity-induced states. 
One fascinating system is a magnetic Weyl semimetal such as the recently discovered 
 Co$_3$Sn$_2$S$_2$, in which the band connecting the Weyl nodes is flattened due to correlations of 3$d$ Co electrons~\cite{Xu_arxiv2019}.

In summary, we showed that flat impurity bands arise in impurity-doped Dirac materials. We  demonstrate this impurity assisted band structure engineering for a specific case of a three-dimensional topological insulator with non-magnetic impurities on the surface. The impurity flat bands hybridize with the Dirac states and modify their nodal structure. Peculiar features, such as the splitting of the Dirac nodes and generation of additional nodes, are predicted and are explained by a low-energy effective model and topological band theory. We argue that these features are not model specific and are not the artefacts of the supercell approach, and are also present in disordered systems. To test the robustness of these predictions in the presence of electron correlation effects we applied the large-$N$ $\textsf{SU}(2)\times \textsf{SU}(N)$ hybridization model to probe the mean field bands and Kondo effect. We find that, while the bands are renormalized, the ``extra" Dirac nodes and flat bands persist in interacting model provided the coupling is sufficiently large.  These results reinforce our predictions of flat bands engineered via impurity bands and render them experimentally feasible.  Predicted impurity-engineered flat bands present a natural platform for exploring interaction-induced phases, beyond the paradigm of twisted bilayer graphene.     

{\em Acknowledgements}  We are grateful to G. Fernando, M. Geilhufe, B. Sinkovic, K. Kaznatcheev,  T. Yilmaz and J.X Zhu for useful discussions. Work was supported by the University of Connecticut, the  European Unions Seventh Framework Program ERC-2018-SyG HERO-810453, by  VILLUM  FONDEN  via  the  Centre  of  Excellence  for  Dirac Materials (Grant No.~11744) and the Knut and Alice Wallenberg Foundation KAW 2019.0068. Work at Brookhaven National Laboratory was supported by the U.S. Department of Energy, Office of Science, Office of Basic Energy Sciences, under Contract No. DESC0012704.
\bibliographystyle{apsrev}
\bibliography{Flat_bands_biblio}

\end{document}